\documentclass[twocolumn,pra,showpacs,amssymb,10pt,superscriptaddress]{revtex4}

\usepackage{graphicx}% Include figure files
\usepackage{dcolumn}% Align table columns on decimal point
\usepackage{bm}% bold math
\usepackage{epsfig}
\usepackage{color}

\def\ga{\mathrel{\mathpalette\fun >}}
\def\fun#1#2{\lower3.6pt\vbox{\baselineskip0pt\lineskip.9pt
\ialign{$\mathsurround=0pt#1\hfil##\hfil$\crcr#2\crcr\sim\crcr}}}
\newcommand{\bfal}{{\mbox{\boldmath$\alpha$}}}

\begin{document}

\preprint{}

%
% ------------------------------ Title --------------------------------------- %
%

\title{Relativistic photoeffect for $s$ states in a central field}

%
% ------------------------------- Authors ------------------------------------ %
%

\author{E.~G.~Drukarev}
\author{A.~I.~Mikhailov}
\affiliation{National Research Center ''Kurchatov Institute'', B.P.~Konstantinov Petersburg Nuclear Physics Institute, Gatchina, 
188300 St.~Petersburg, Russia}

\author{Kh.~Yu.~Rakhimov}
\affiliation{Department of Chemistry, University of Antwerp, Universiteitsplein 1, 2610 Antwerpen, Belgium}
%\affiliation{Department of Physics, National University of Uzbekistan, 100174 Tashkent, Uzbekistan}

\author{H.~T.~Yusupov}
\affiliation{Faculty of Physics and Mathematics, Tashkent State Pedagogical University, 27 Bunyodkor Str., Tashkent 100070, Uzbekistan}

\date{\today \\[0.3cm]}

%
% ------------------------------- Abstract ----------------------------------- %
%

%
%
%
\begin{abstract}
We study the photoionization of the $s$ states in the systems bound by sufficiently weak central fields $V(r)$ for the large photon energies
corresponding to the relativistic photoelectrons. We demonstrate that the energy dependence of the photoionization cross section can be obtained without solving the wave equation. We show that the shape of the energy dependence of the cross section is determined by analytical properties of the binding potential $V(r)$.  We find the cross sections for the potentials $V(r)$ which have singularities in the origin, on the real axis and in the complex plane.
\end{abstract}

 \maketitle

\section{Introduction}

We study the ionization of systems bound by a central field $V(r)$ by the photons carrying the energy $\omega$ of the order of the electron rest energy $mc^2$. In the present paper we include only photoionization of the single particle bound states with the orbital momenta $\ell=0$.
We consider the lowest order terms in the coupling constant $g$ of the field $V(r)$.
We find the general expressions for the distribution in recoil momenta and for the total cross sections.

The characteristics of the relativistic photoeffect on $1s$ states in the Coulomb field in the lowest order of the coupling constant $g=\alpha Z$ ($\alpha=1/137$ is the fine structure constant, $Z$ is the charge of the nucleus) were found in~\cite{Sauter} in the early days of quantum mechanics. Further progress was made by considering the higher terms in $\alpha Z$ expansion for the Coulomb field~\cite{Gavrila,Weber,Gorshkov}. Numerical calculations for the screened Coulomb field were also carried out~\cite{Pratt}. The analysis of~\cite{Sauter,Gavrila,Weber,Gorshkov} as well as the presentation in the books~\cite{Zommerfeld,Berestetskii} employ the explicit form of the Coulomb $1s$ bound state wave function as a starting point.

However expressions for the angular distributions of photoelectrons or in recoil momenta and for the total cross sections can be obtained for a
broad class of central potentials  $V(r)$. The basic point of our approach is the analysis of the process in terms of three-momentum transferred to the electrons by the source of the field
\begin{equation}
{\bf q}={\bf p}-{\bf k}.
\label{0}
\end{equation}
Here ${\bf k}$ and ${\bf p}$ are the three-momenta of the photon and of photoelectron correspondingly. The source of the field obtains the recoil momentum $-{\bf q}$. We presented the approach in our book~\cite{Drukarev} and applied it later to investigation of the high energy nonrelativistic asymptotics of photoionization~\cite{Drukarev1,Drukarev2}.

We employ the system of units with $\hbar=1$, $c=1$. In these units the photon three momentum ${\bf k}$ and its energy are related by the condition $\omega=k \equiv |{\bf k}|$. The energy conservation law is
\begin{equation}
E_B+\omega=E,
\label{0n}
\end{equation}
with $E_B=m-I_B$ and $E=\sqrt{p^2+m^2}$ the relativistic energies of the bound electron and of the photoelectron.
We consider the photon energies $\omega \ga m$ with $m \approx 511$ keV the electron mass.

We assume that the electron binding energy $I_B$ is much smaller than the electron mass, i.e. $I_B \ll m$. Thus we have two scales of momenta. The characteristic momenta exchanged between the bound electron and the nucleus are of the order $\mu=(2mI_B)^{1/2}$. Since $I_B \ll m$ we find that
\begin{equation}
\mu \ll m.
\label{0n1}
\end{equation}
We call it "small momenta". The value of transferred  momentum is limited by conditions
\begin{equation}
p-k \leq q \leq p+k,
\label{0a}
\end{equation}
and thus $q \ga m$.
We call it "large momenta".

In Sec.~2 we demonstrate that if the Fourier transform $\tilde V(q)$ (we omit the tilde sign below) becomes small enough at $q \ga m$
\begin{equation}
\frac{q^2|V(q)|}{2\pi^2} \ll 1,
\label {0b}
\end{equation}
each act of transfer of large momentum leads to additional small factor in the amplitude (see Chapter~2 of~\cite{Drukarev}). Thus such exchange can be treated perturbatively. This enables to obtain unified expression for the recoil momenta distributions and for the cross sections presented in Sec.~3.

We demonstrate that the $q$ dependence of the recoil momentum distributions can be presented through the Fourier transform $V(q)$ of the potential $V(r)$. The shape of the potential $V(q)$ is determined by the singularities of the function $V(r)$. Thus the recoil momentum distributions and the total cross sections are determined by singularities of the potential $V(r)$.

In Sec.~4-6 we consider the potentials with singularities of the potential $V(r)$ at the origin, on the real axis and in the complex plane correspondingly. We summarize in Sec.~7.

The photoionization cross section of a bound state with one electron can be written as~\cite{Berestetskii}
\begin{equation}
d\sigma=pE|\bar F|^2\delta(E-\omega-m+I_B)\frac{dEd\Omega}{4\pi^2}=pE|\bar F|^2\frac{d\Omega}{4\pi^2},
\label{1}
\end{equation}
with $\Omega$ the photoelectron solid angle.
The overbar indicates that the squared amplitude $F^2$ is averaged over polarizations in the initial state and summed over those in the final state.

We start with calculation of the amplitude.

\section{Photoionization amplitude}

The amplitude is
\begin{equation}
F=(4\pi\alpha)^{1/2}\langle\Psi_{\bf p}|-A_{i}\gamma^{i}|\Psi_B\rangle; \quad i=1,2,3.
\label{2}
\end{equation}
Here $\gamma^{\mu}$ are the  Dirac matrices, $A_i$ are the space components of the four-vector $A_{\mu}$ describing the electromagnetic field.
The relativistic wave functions $\Psi_{B, {\bf p}}$ describe the bound electron and the photoelectron with asymptotic three momentum ${\bf p}$. They satisfy the Dirac equations
\begin{equation}
(H_0+V(r))\Psi_B=E_B\psi_B; \quad (H_0+V(r))\Psi_{\bf p}=E\Psi_{\bf p},
\label{2a}
\end{equation}
with $E_B=m-I_B$ and  $E=(p^2+m^2)^{1/2}$. In Eq.~(\ref{2a})
\begin{equation}
H_0=\alpha_{i}f^i+\beta m,
\label{2c}
\end{equation}
is the Hamiltonian of free motion. Here $\alpha_i=\gamma_0\gamma_i$ ($i=1,2,3$), $\beta=\gamma_0$. In position space ${\bf f}=-i{\bf\nabla}$.
Eq.~(\ref{2a}) can be written as
\begin{equation}
(\hat f-V(r)\gamma_0)\Psi_{B,{\bf p}}=m\Psi_{B, {\bf p}}.
\label{2d}
\end{equation}
We denote $\hat a=a_{\mu}\gamma^{\mu}=a^{\mu}\gamma_{\mu}$ for any four vector $a=(a_0,{\bf a})$. In Eq.~(\ref{2d}), $f_0=m-I_B$ for the bound electron, while $f_0=(p^2+m^2)^{1/2}$ for the photoelectron. The equation for free motion with the energy $f_0$ is
\begin{equation}
(\hat f-m)\Psi^{(0)}=0.
\label{2e}
\end{equation}

We carry out most of calculations in momentum space, where $A_i=e_i/\sqrt{2\omega}$, while ${\bf e}$ is the photon polarization vector, ${\bf e}{\bf k}=0$.
The amplitude can be written as
\begin{equation}
F=N(\omega)\int\frac{d^3s}{(2\pi)^3}\bar\Psi_{\bf p}({\bf k}+{\bf s})(-e_i\gamma^i)\Psi_B({\bf s}),
\label{5}
\end{equation}
where $N(\omega)=\sqrt{4\pi\alpha/2\omega}$. The Eq.~(\ref{5}) can be obtained by Fourier transform of the wave functions in more familiar space presentation where
$A_i=e_ie^{i{\bf k}{\bf r}}/\sqrt{2\omega}$ and
$$F=N(\omega)\int d^3r\bar\Psi_{\bf p}({\bf r})(-e_i\gamma^i)e^{i{\bf k}{\bf r}}\Psi_B({\bf r}).$$

It is reasonable to consider separately the configuration in which the photoelectron does not transfer any momentum to the source of the field.
The contribution to the amplitude is
\begin{equation}
F_a=N(\omega)\int\frac{d^3s}{(2\pi)^3}\bar\Psi^{(0)}_{\bf p}({\bf k}+{\bf s})(-e_i\gamma^i)\Psi_B({\bf s}).
\label{6}
\end{equation}
Here $\Psi^{(0)}_{\bf p}({\bf k}+{\bf s})=\delta({\bf p}-({\bf k}+{\bf s}))u_p=\delta({\bf q}-{\bf s})u_p$
is the solution of the wave equation (\ref{2e}) for free motion
with $f_0=(p^2+m^2)^{1/2}$. The bispinor $u_p=u(E,{\bf p})$ is normalized by condition $\bar uu=2m$.
Thus ${\bf s}={\bf q}$ in the integrand on the right hand side of Eq.~(\ref{6}), and
\begin{equation}
F_a=N(\omega)\bar u_p(-e_i\gamma^i)\Psi_B({\bf q}).
\label{7}
\end{equation}

To include the possibility of photon exchanges between the photoelectron and the source of the field introduce the  functions $\Phi_{\bf  p}= \Psi_{\bf p}-\Psi^{(0)}_{\bf p}$. The corresponding contribution to the amplitude is
\begin{equation}
F_b=N(\omega)\int\frac{d^3s}{(2\pi)^3}\bar \Phi_{\bf p}({\bf k}+{\bf s})(-e_i\gamma^i)\Psi_B({\bf s}),
\label{8}
\end{equation}
and
\begin{equation}
F=F_a+F_b.
\label{9}
\end{equation}
Now we calculate wave functions $\Psi_B$ and $\Phi_{\bf p}$ at large values of the arguments.

We can write for any electron state with the energy $f_0$
\begin{equation}
\Psi=\Psi^{(0)}+G_0(f_0)V\gamma_0\Psi,
\label{10}
\end{equation}
with
\begin{equation}
G_0(f_0)=(\hat f-m)^{-1},
\label{10a}
\end{equation}
the Green function of the free motion equation (\ref{2e}). One can obtain Eq.~(\ref{10a}) by subtracting Eqs.~(\ref{2d}) and ({\ref{2e}) at the same value of $f_0$.

For the bound state wave function $\Psi^{(0)}_{B}=0$. Hence Eq.~(\ref{10}) can be written as
\begin{equation}
\Psi_B({\bf q})=\int\frac{d^3h}{(2\pi)^3}\frac{d^3 f}{(2\pi)^3}\langle{\bf q}|G_0(m-I_B)|{\bf h}\rangle
\langle{\bf h}|V|{\bf f}|\rangle\gamma_0 \Psi_B({\bf f}).
\label{11}
\end{equation}
Note that the matrix element of the Green function with the energy ${\cal E}$ is
\begin{equation}
\langle{\bf q}|G_0({\cal E})|{\bf h}\rangle=\frac{\hat h+m}{h^2-m^2+i\delta}\delta({\bf h}-{\bf q}), \quad \delta \rightarrow 0.
\label{12}
\end{equation}
with the four-vector $h=({\cal E},{\bf h})$.
Employing this equality to the matrix element in the integrand on the right hand side of Eq.~(\ref{11}) with $f_0=m-I_B$ and carrying out integration over ${\bf h}$ we find
\begin{equation}
\Psi_B({\bf q})=-\frac{m+m\gamma_0+q^{i}{\alpha}_{i}}{q^2}\cdot J_{rel}({\bf q}),
\label{14}
\end{equation}
with
\begin{equation}
J_{rel}({\bf q})=\int\frac{d^3f}{(2\pi)^3}\langle{\bf q}|V|{\bf f}\rangle\Psi_B({\bf f}),
\label{14a}
\end{equation}
and $\alpha_i=\gamma_0\gamma_i$.

One can separate three regions of the values of momentum $f$ in the integral $J_{rel}(q)$. We analyze their contributions to wave function determined by Eq.~(\ref{14}). In the region $f \sim \mu \ll q$, we can estimate $V({\bf q}-{\bf f})=V({\bf q})$.
Hence it provides a contribution of the order $\mu^3V(q)\Psi_B(\mu)/2\pi^2$ to the integral $J_{rel}(q)$ and contribution $\mu^3V(q)\Psi_B(\mu)/(2\pi^2q)$ to the wave function $\Psi_B({\bf q})$. The  large momenta $f \sim q$ for which $|{\bf q}-{\bf f}| \sim \mu \ll q$  provide a contribution of the order $\mu^3V(\mu)\Psi_B(q)/2\pi^2$ to  $J_{rel}(q)$. This causes a small correction of the order $\mu/q$ to the function $\Psi_B(q)$ which can be neglected.
Similar estimation of contribution of the region of large momenta $f \sim q$ ; $|{\bf q}-{\bf f}| \sim q$ shows that it
provides a correction $V(q)q^2/2\pi^2$ to the function $\Psi_B$ if Eq.~(\ref{0b}) is true. In this case it can be neglected.
Hence the integral $J_{rel} (q)$ is saturated by small $f \sim \mu$ for the potentials which satisfy the condition expressed by Eq.~(\ref{0b}).

At $f \sim \mu \ll m$ the bound state wave function can be expressed as
\begin{equation}
\Psi_B({\bf f})=\psi_B(f)u_0.
\label{15a}
\end{equation}
with $\psi_B(f)$ the nonrelativistic bound state wave function (it does not depend on direction of momentum ${\bf f}$), while $u_0=u(m,{\bf f}=0)$.
Hence we can write
\begin{equation}
J_{rel}(q)=J(q)u_0,
\label{16b}
\end{equation}
with
\begin{equation}
J(q)=\int\frac{d^3f}{(2\pi)^3}V({\bf q}-{\bf f})\psi_B(f),
\label{16b1}
\end{equation}
which also does not depend on direction of momentum ${\bf q}$.
Employing the equality $\gamma_0u_0=u_0$ we can write
\begin{equation}
\Psi_B({\bf q})=-\frac{2m}{q^2}(1+\frac{\bfal \bf q}{2m})J(q)u_0.
\label{16b2}
\end{equation}
 Thus Eqs.~(\ref{16b1}) and (\ref{16b2}) connect relativistic wave function $\Psi_B({\bf q})$ at $q \ga m$ with nonrelativistic function $\psi_B(f)$ at $f \sim \mu$.

Thus
\begin{equation}
F_a=-N(\omega)J(q)\bar u_p\frac{\hat e(2m+\tilde q)}{q^2}u_0,
\label{16a}
\end{equation}
with $\hat e=-e_i\gamma^i$, $\tilde q=q_i\alpha^i$.

To calculate the contribution $F_b$ to the amplitude we need to find the function $\Phi_{\bf p}({\bf k}+{\bf s})$. We employ the first iteration of Eq.~(\ref{10})
\begin{equation}
\Psi_{\bf p}=\Psi^{(0)}_{\bf p}+G_0(m+\omega-I_B)V\gamma_0\Psi_{\bf p},
\label{17}
\end{equation}
i.e.
\begin{equation}
\Phi_{\bf p}=G_0(m+\omega-I_B)V\gamma_0\Psi^{(0)}_{\bf p}.
\label{18}
\end{equation}
Thus
$$
\Phi_{\bf p}({\bf k}+{\bf s})=\int\frac{d^3f}{(2\pi)^3}\frac{d^3h}{(2\pi)^3}\langle{\bf k}+{\bf s}|G_0(m-I_B+\omega)|{\bf h}\rangle
$$
\begin{equation}
\times\langle{\bf h}|V|{\bf f}\rangle\gamma_0 \Psi^{(0)}_{\bf p}({\bf f}).
\label{19}
\end{equation}
Carrying out integrations and denoting ${\bf k}'={\bf k}+{\bf s}$ we obtain
\begin{equation}
\Phi_{\bf p}({\bf k}')=\frac{\hat P+m}{P^2-m^2+i\delta}V({\bf p}-{\bf k}')\gamma_0u_p; \quad P=(E, {\bf k}'),
\label{20}
\end{equation}
with $E=\omega+m$ the relativistic energy of the photoelectron. We neglected $I_B \ll m$ in expression for $E$. As we have seen, for the potential which satisfy the condition expressed by Eq.~(\ref{0b})
integral on the right hand side of Eq.~(\ref{8}) will be saturated by small $s \sim \mu$. Thus in the integrand of Eq.~(\ref{8}) we can use the function $\Phi_{\bf p}({\bf k}+{\bf s})$ in which  we put $P=(E, {\bf k})$ and $P^2=2m\omega+m^2$, i.e
\begin{equation}
\Phi_{\bf p}({\bf k}+{\bf s})=\frac{\hat P+m}{2m\omega}V({\bf q}-{\bf s})\gamma_0u_p.
\label{21}
\end{equation}
We can evaluate $(\hat P+m)\gamma_0u_p=(\hat p-\hat q +m)\gamma_0u_p$ with the four vector $q=(0,{\bf q})$. Hence $(\hat P+m)\gamma_0u_p=(2E-\tilde q)u_p$. The last equality is due to relation $\hat p\gamma_0u_p=
(2E-m\gamma_0)u_p$.
Thus
\begin{equation}
F_b=N(\omega)J(q)\bar u_p\frac{(2E+\tilde q)\hat e}{2m\omega}u_0.
\label{23a}
\end{equation}
Recall that the amplitude $F=F_a+F_b$.

\section{Differential distribution and the cross section}
Employing Eqs.~(\ref{9}), (\ref{16a}), (\ref{23a}), and Eq.~(\ref{1}) we find the angular distribution of photoelectrons
\begin{equation}
\frac{d\sigma}{dt}=\alpha\frac{p}{\omega}J^2(q)W(t),
\label{24}
\end{equation}
with
$t=\cos{\theta}$, while $\theta$ is the angle between the directions of momenta ${\bf k}$ and ${\bf p}$,
\begin{equation}
W(t)=\frac{2mp^2(1-t^2)}{q^4}\Big(1+\frac{q^2}{4m^2}(\frac{\omega}{m}-1)\Big).
\label{25}
\end{equation}
Here $q^2=q^2(t)=p^2+\omega^2-2p\omega t=2\omega(m+\omega-pt)$.

Distribution over the recoil momentum is
\begin{equation}
\frac{d\sigma}{qdq}=\alpha\frac{J^2(q)}{\omega^2}W(q),
\label{26}
\end{equation}
where $W(q)$ is determined by Eq.~({\ref{25}) with
\begin{equation}
t(q)=\frac{E}{p}-\frac{q^2}{2\omega p}.
\label{27}
\end{equation}

Thus
\begin{equation}
\frac{d\sigma}{dq}=\alpha T(q),
\label{26a}
\end{equation}
with
\begin{equation}
T(q)=\sum_{n=0}^3C_nA_n(q)
\label{26b}
\end{equation}
where
\begin{equation}
C_0=-\frac{(E-2m)}{8m^2(E-m)^4};\quad C_1=\frac{(E^3-3E^2m+2Em^2-m^3)}{2m^2(E-m)^4},
\label{26c}
\end{equation}
$$C_2=-\frac{(E^2-7Em+2m^2)}{2(E-m)^3}; \quad C_3=-\frac{2m^3}{(E-m)^2}.$$
Dependence on the potential $V$ is contained in the factors
\begin{equation}
A_n(q)=\frac{J^2(q)}{q^{2n-3}}.
\label{26d}
\end{equation}

Thus the cross section is
\begin{equation}
\sigma_V(E)=\alpha \int_{q_1}^{q_2}T_V(q)dq,
\label{26e}
\end{equation}
with
\begin{equation}
q_1=p-\omega=\sqrt{E^2-m^2}-(E-m),\\
\label{26g}
\end{equation}
$$
q_2=p+\omega=\sqrt{E^2-m^2}+(E-m),
$$
are the smallest and the largest possible values of the residual momentum $q$ correspondingly.
Lower index $V$ in Eq.~(\ref{26e}) labels the type of the potential.

Now we consider the potentials $V(r)$ with various positions of singularities. We start with potentials with singularities at the origin $r=0$. They appear to be "slowly varying potentials" for which
\begin{equation}
\mu V'(q) \ll V(q).
\label {0c}
\end{equation}
This enables us to expand $V({\bf q}-{\bf f})$ in Eq.~(\ref{16b}) for $J(q)$ in powers of $f/q$ putting
$V({\bf q}-{\bf f})=V(q)$. This provides
\begin{equation}
J(q)=V(q)\psi_B(r=0),
\label{27a}
\end{equation}
and the recoil momentum distribution is indeed determined by the $q$ dependence of the potential $Vq)$.
We shall see that the same refers to the potentials with singularities on the real axis and in the complex plane although neither Eq.~(\ref{0c}) nor ({\ref{27a}) are true.

Now we analyze particular cases.

\section{Potentials with singularity at the origin}
\subsection{Coulomb potential}

The Fourier transform for the Coulomb potential of the nucleus containing $Z$ protons $V_C(r)=-\alpha Z/r$ is
\begin{equation}
V_C(q)=-\frac{4\pi \alpha Z}{q^2}.
\label{28}
\end{equation}
For any bound $ns$ state
\begin{equation}
J(q)=-\frac{4\pi \alpha Z}{q^2}\psi_C(r=0),
\label{29}
\end{equation}
with $\psi^2_C(r=0)=\eta ^3_n/\pi$, $\eta_n=m\alpha Z/n$.
The characteristic binding momentum of $ns$ state is $\mu=\eta_n=m\alpha Z/n$.

Employing Eqs.~(\ref{26b})-(\ref{26e}) we find for photoionization of $ns$ state
\begin{equation}
\sigma_C(\omega)=2\alpha(\alpha Z)^2\pi^2\frac{\psi_C^2(r=0)}{m^5}S_C(\zeta),
\label{30}
\end{equation}
with $\zeta=m/E$, and
\begin{equation}
S_C(\zeta)=\frac{\zeta^2(1-\zeta^2)^{3/2}}{(1-\zeta)^5}\cdot
\left(\frac{4}{3}+\right.
\label{30a}
\end{equation}
$$
+\frac{1-2\zeta}{\zeta(1+\zeta)}\left.\Big(1-\frac{\zeta^2}{2\sqrt{1-\zeta^2}}
\ln{\frac{1+\sqrt{1-\zeta^2}}
{1-\sqrt{1-\zeta^2}}}\Big)\right),
$$
%see Fig.1.
see Fig.~\ref{fig:Fig1}.
This expression was obtained in~\cite{Sauter} for $1s$ electrons.

\begin{figure}[t]
\begin{center}
\includegraphics[width=7cm]{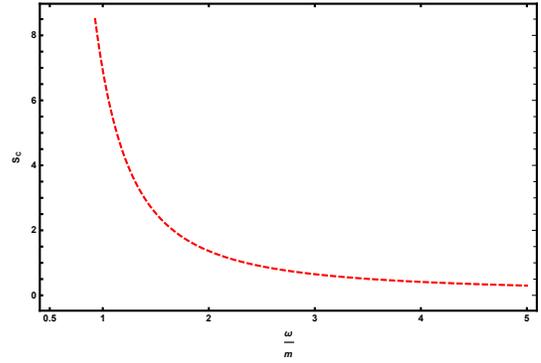}
\caption{Energy dependence for photoionization in the case of the Coulomb field. The horizontal axis is for the ratio
 $\omega/m$ with $\omega$ the photon energy while $m \approx 511$ keV is the electron mass. The vertical axis is for the function $S_C$ given by Eq.~(49).}
\label{fig:Fig1}
\end{center}
\end{figure}

% Fig. 1. Energy dependence for photoionization in the case of the Coulomb field. The horizontal axis is for the ratio
% $\omega/m$ with $\omega$ the photon energy while $m \approx 511$ keV is the electron mass. The vertical axis is for the function $S_C$ given by Eq.(49).\\

In the ultrarelativistic limit $\zeta \ll 1$ we find $S_C(\zeta)=\zeta$, and
\begin{equation}
\sigma_C(\omega)=2\alpha(\alpha Z)^2\pi^2\frac{\psi_C^2(r=0)}{m^5}\frac{m}{\omega}.
\label{30x}
\end{equation}

There are two characteristic parameters of the motion in the Coulomb field. Average velocity of the electron in the ground state is $\alpha Z$.
Interaction of continuum electron with the nucleus is described by the parameter $\xi=\alpha Z E/p$. The distribution provided by Eq.~(\ref{26a}) and the cross section given by Eq.~(\ref{30}) correspond to the lowest order expansions in both of them. However the leading $\xi$ dependent contributions depend on parameter $\pi\xi$. Thus the coefficients of expansion in $\xi$ become numerically large. Fortunately the dependence on $\pi\xi$ (as well as in nonrelativistic case - see~\cite{Drukarev}) can be calculated accurately~\cite{Gorshkov}. It is expressed by the function
\begin{equation}
{\cal S}(\xi)=\frac{2\pi\xi e^{-2\pi\xi}}{1-e^{-2\pi\xi}}=e^{-\pi\xi}|\Gamma(1+i\xi)|^2=e^{-\pi\xi}\Big(1+0(\xi^2)\Big),
\label{31}
\end{equation}
known as the Stobbe factor. Thus
\begin{equation}
\frac{d\sigma_C^{LO}}{dq}=\frac{{\cal S}(\xi)d\sigma_C}{dq}; \quad  \sigma_C^{LO}(\omega)={\cal S}(\xi)\sigma_C(\omega).
\label{32}
\end{equation}
Here the upper index $LO$ shows that the leading order corrections which depend on $\pi\xi$ are included.

\subsection{Screened Coulomb potential}

Consider now the screened Coulomb potential
\begin{equation}
V_{scr}(r)=-\alpha Z\frac {f(\lambda r)}{r}; \quad f(0)=1,
\label{33}
\end{equation}
with $\lambda \sim m\alpha Z $.

The Yukawa potential
\begin{equation}
V_{Y}(r)=-\alpha Z\frac {e^{-\lambda r}}{r}; \quad V_{Y}(q)=-\frac{4\pi\alpha Z}{q^2+\lambda^2},
\label{34}
\end{equation}
provides an example. We find immediately that $V_Y(q)=-4\pi\alpha Z/q^2=V_C(q)$ for $q \ga m$.
Thus the distribution in recoil momentum and the cross section are expressed by Eq.~(\ref{26a}) and Eq.~(\ref{30}) with the Coulomb value $\psi_C^2(r=0)$ (see Eq.~(\ref{29}) replaced by those of the Yukawa value $\psi_Y^2(r=0)$.

One can make a more general statement. The Fourier transform of the screened potential (\ref{33}) at $q \ga m$ coincides with that of the Coulomb potential
with the accuracy $\lambda^2/q^2$
\begin{equation}
V_{scr}(q)=-\frac{4\pi\alpha Z}{q^2}\Big(1+0(\lambda^2/q^2)\Big).
\label{35}
\end{equation}
This was demonstrated in~\cite{Drukarev} by expansion of the function $f(\lambda r)$ in the right hand side of (\ref{33}) in powers of $\lambda r$.
Thus the characteristics of the process in the lowest order of $\alpha Z$ expansion $d\sigma_{scr}/dq$ and $\sigma_{scr}$ are described by the same equations as in the Coulomb field with $\psi^2_{C}(r=0)$ replaced
by $\psi^2_{scr}(r=0)$.

It was shown in~\cite{Drukarev2} that the shape of high energy nonrelativistic photoionization cross section of manyelectron atoms is reproduced by inclusion of the Stobbe factor. This happens because the latter is formed at small distances $1/p$ from the nucleus where the screening effects are small. One can expect that this is true also for relativistic photoelectrons, and
\begin{equation}
\frac{d\sigma_{scr}^{LO}}{dq}=\frac{{\cal S}(\xi)d\sigma_{scr}}{dq}; \quad  \sigma^{LO}_{scr}(\omega)={\cal S}(\xi)\sigma_{scr}(\omega).
\label{36}
\end{equation}

Thus we obtained
\begin{equation}
\frac{\sigma_C(\omega)}{\sigma_{scr}(\omega)}=\frac{\psi_{C}^2(r=0)}{\psi_{scr}^2(r=0)},
\label{37}
\end{equation}
with similar relation for the cross sections which include the leading corrections $\sim \pi\xi$. The numerical calculations~\cite{Pratt} demonstrated that Eq.~(\ref{37}) is true with the accuracy of about $1$ percent.

\subsection {Exponential potential}

The exponential potential is
\begin{equation}
V_{Exp}(r)=-V_0e^{-\lambda r} ; \quad V_0>0.
\label{38}
\end{equation}
Considered as a function of $-\infty <r<\infty$ it can be written as $V_{Exp}(r)=-V_0e^{-\lambda |r|}$ with a cusp at $r=0$. The exponential potential found its applications in cosmology - see, e.g.~\cite{Ratra}.
Its Fourier transform can be obtained immediately by taking the derivative with respect to $\lambda$ of the equations  on right hand sides of Eq.~({\ref{34})
\begin{equation}
V_{Exp}(q)=-\frac{8\pi\lambda V_0}{q^4}.
\label{39}
\end{equation}
We consider the case $V_0\lambda/m^2 \ll 1$ - see Eq.~(4).
For any bound $s$ state
\begin{equation}
J(q)=-\frac{8\pi\lambda V_0}{q^4}\psi_{Exp}(r=0).
\label{40}
\end{equation}

Employing Eqs.~(\ref{26b})-(\ref{26e})
we find for the cross section
\begin{equation}
\sigma_{Exp}(\omega)=\frac{4}{3}\alpha \pi^2\Big(\frac{V_0\lambda}{m^2}\Big)^2\frac{\psi_{Exp}^2(r=0)}{m^5}S_{Exp}(\zeta); \quad \zeta=\frac{m}{E},
\label{41}
\end{equation}
with  $\psi_E$ the wave function of $s$ bound state in exponential potential (\ref{38}) and
\begin{equation}
S_{Exp}(\zeta)=\frac{\zeta(1-3\zeta/5+2\zeta^2-2\zeta^3)(1-\zeta^2)^{3/2}}{(1-\zeta)^7},
\label{41a}
\end{equation}
see Fig.~\ref{fig:Fig2}. At $E \gg m$, i.e. at $\zeta \ll 1$, we find $S_{Exp}(\zeta)=\zeta$, and
\begin{equation}
\sigma_{Exp}(\omega)=\frac{4}{3}\alpha \pi^2\Big(\frac{V_0\lambda}{m^2}\Big)^2\frac{\psi_{Exp}^2(r=0)}{m^5}\cdot\frac{m}{\omega}.
\label{41b}
\end{equation}

\begin{figure}
\begin{center}
\includegraphics[width=7cm]{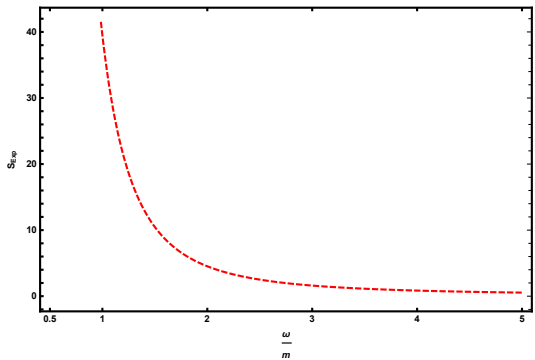}
\caption{Energy dependence for photoionization in the case of the exponential potential defined by Eq.~(58). The horizontal axis is the same as in Fig.~\ref{fig:Fig1}. The vertical axis is for the function $S_{Exp}$ given by Eq.~(62).
}
\label{fig:Fig2}
\end{center}
\end{figure}

% Fig. 2. Energy dependence for photoionization in the case of the exponential potential defined by Eq.(58). The horizontal axis is the same as in Fig.1. The vertical axis %is for the function $S_{Exp}$ given by Eq.(62).\\

\section{Potentials with singularities on the real axis}

Start with  potential which has a finite jump on the real axis, i.e.
$V(r)=V_1(r)$ at $r\leq R$ and $V(r)= V_2(r)$ at $R<r<\infty$ with $V_1(R)\neq V_2(R)$. This can be expressed as
\begin{equation}
V_{FJ}(r)=V_1(r)\theta(R-r)+V_2(r)\theta(r-R).
\label{60}
\end{equation}
The lower index $FJ$ stands for "finite jump".
Recall that $\theta(x)=1$ for $x \geq 0$ while $\theta(x)=0$ for $x<0$.

We assume $R$ to be of the order of characteristic atomic size. Thus
\begin{equation}
mR \gg 1.
\label{61}
\end{equation}
We consider the potentials which, as well as their derivatives, change on the distances of the atomic size $r_B \sim 1/\mu$ - see Eq.~(\ref{0n1}). The relation between the two characteristics with dimension of length $R$ and $r_B$ is not important for us since $(mR)^{-1} \ll 1$ and $(mr_B)^{-1} \ll 1$, and we include only the lowest terms of expansions in these parameters.

The Fourier transform of the potential (\ref{60}) is
\begin{equation}
V_{FJ}(q)=\int d^3re^{-i {\bf q}{\bf r}}V_{FJ}(r)=
\label{62}
\end{equation}
$$\frac{4\pi}{q}\lim|_{\delta \rightarrow 0}\Big(\int_0^{R-\delta} drr\sin{(qr)}V_{FJ}(r)$$
$$+\int_{R+\delta}^{\infty}drr\sin{(qr)}V_{FJ}(r)\Big).$$
Integration by parts provides
\begin{equation}
V_{FJ}(q)=\frac{4\pi}{q}\Big(\frac{ \cos{(qR)}\cdot \delta_V(R)}{q}
\label{63}
\end{equation}
$$+\lim_{\delta \rightarrow 0}[\frac{1}{q}\int_0^{R-\delta} dr\cos{(qr)}V'(r)+\frac{1}{q}\int_{R+\delta}^{\infty}\cos{(qR)}V'(r)]\Big),$$
Successive integration by parts on the right hand side of Eq.~({\ref{63}) provides series in powers of
$(qr_B)^{-1}\ll 1$ and $(qR)^{-1}\ll 1$. Thus we include only the first term in big parenthesis and
\begin{equation}
V_{FJ}(q)=4\pi\delta_V(R)\frac{\cos{(qR)}}{q^2},
\label{63n}
\end{equation}
with
\begin{equation}
\delta_V(R)=\lim_{\delta \rightarrow 0}\Big (V(R+\delta)-V(R-\delta)\Big).
\label{63a}
\end{equation}

We can put $({\bf q}-{\bf f})^2=q^2$ in the denominator of expression for the potential $V({\bf q}-{\bf f})$ on the right hand side of Eq.~(\ref{16b}) since $f \sim \mu \ll q$. The factor $\cos{(|{\bf q}-{\bf f}|R)}$
can not be expanded in powers of $f/q$, since $R \sim 1/\mu$ and thus $fR \sim 1$. However we can put $|{\bf q}-{\bf f}|R=(q-f_q)R$ with
$f_q={\bf q}{\bf f}/q$, and find that
\begin{equation}
J({\bf q}-{\bf f})=V_{FJ}(q)\int\frac{d^3f}{(2\pi)^3}\cos{(f_qR)}\psi_{FJ}(f).
\label{65n1}
\end{equation}
Thus the $q$ dependence of the factor $J(q)$ defined by Eq.~(\ref{16b1}) is determined by that of the potential $V_{FJ}(q)$.

Expression for $J(q)$ can be obtained by evaluation in the position space.
We present
\begin{equation}
J(q)=\int d^3re^{-i {\bf q}{\bf r}}V(r)\psi_{FJ}(r)=
\label{62x}
\end{equation}
$$\frac{4\pi}{q}\lim|_{\delta \rightarrow 0}\Big(\int_0^{R-\delta} dr\sin{(qr)}F(r)+\int_{R+\delta}^{\infty}dr\sin{(qr)}F(r)\Big).$$
Here $F(r)=r\psi_{FJ}(r)V(r)$.
Integration by parts provides
\begin{equation}
J(q)=\frac{4\pi}{q}\Big(\frac{\cos{(qR)}\cdot \delta_F(R)}{q}
\label{63nn}
\end{equation}
$$+\lim_{\delta \rightarrow 0}[\frac{1}{q}\int_0^{R-\delta} dr\cos{(qr)}F'(r)+\frac{1}{q}\int_{R+\delta}^{\infty}dr\cos{(qR)}F'(r)]\Big),$$
with $\delta_F(R)=\lim_{\delta \rightarrow 0}\Big (F(R+\delta)-F(R-\delta)\Big)$.
Since the wave function $\psi_{FJ}(r)$ is continuous at $r=R$~\cite{Schiff} we find $\delta_F(R)=R\psi(R)\delta_V(R)$ with $\delta_V(R)$ provided by Eq.~(\ref{63a}). Proceeding in the same way as we did above in calculation of $V(q)$ we obtain
\begin{equation}
J(q)=4\pi R\psi_{FJ}(R)\delta_V(R)\frac{\cos{(qR)}}{q^2},
\label{65}
\end{equation}
with $\delta_V(R)$ defined by Eq.~(\ref{63a}).

Recall that the dependence of the cross section on the form of the potential is contained in the factors $J^2(q)$ (see Eqs.~(\ref{26a})-({\ref{26e})). We can write
\begin{equation}
J^2(q)={(4\pi R \delta_V(R)\psi_{FJ}(R))^2}\frac{\cos^2({qR})}{q^4}.
\label{66}
\end{equation}
Employing the well known relation $\cos^2{\varphi}=(1+\cos{(2\varphi)})/2$, we present $J^2(q)=J^{2(1)}(q)+J^{2(2)}(q)$ with
$$
J^{2(1)}(q)=\frac{(4\pi R \delta_V(R)\psi_{FJ}(R))^2}{2q^4},
$$
\begin{equation}
J^{2(2)}(q)=J^{2(1)}(q)\cos{(2qR)}.
\label{67}
\end{equation}

The integrand in Eq.~(\ref{26e}) presenting the cross section is the sum of the two terms $T_{FJ}^{(1)}(q)$ and $T_{FJ}^{(2)}(q)$ containing the factors
$J^{2(1)}(q)$ and $J^{2(2)}(q)$ correspondingly. The latter contains additional strongly oscillating factor
$\cos{(2qR)}$ with $qR \gg 1$. Thus we can expect that contribution to the cross section caused by $T_{FJ}^{(2)}(q)$ is much smaller than that provided by $T_{FJ}^{(1)}(q)$. This is supported by the following calculation.
Note that $T_{FJ}^{(1)}(q)$ is a composition of the terms $\int_{q_1}^{q_2}dq/{q^k}\sim 1/(q_i)^{k-1}$ ($i=1,2)$.
The corresponding terms in $T_{FJ}^{(2)}(q)$ are
$\int_{q_1}^{q_2}dq \cos{(2qR)}/q^k=(2R)^{k-1}X(R)$,
with
$$X(R)=\int_{2q_1R}^{2q_2R}dy\frac{\cos{y}}{y^k}=
\frac{\sin{2q_2R}}{(q_2R)^k}
$$
$$-\frac{\sin{2q_1R}}{(q_1R)^k}
+k\int_{2q_1R}^{2q_2R}dy\frac{\sin{y}}{y^{k+1}}.$$
Further integration by parts provides a series in $(q_iR)^{-1}\ll 1$. Thus the contribution caused by
$J^{2(2)}(q)$ to the cross section is at least $(q_iR)^{-1}$ times smaller than that of $J^{2(1)}(q)$.
Hence in calculation of the total cross section we can put
\begin{equation}
J^2(q)=J^{2(1)}(q)=\frac{(4\pi R \delta_V(R)\psi_{FJ}(R))^2}{2q^4},
\label{68}
\end{equation}
with the same $q$ dependence as for the Coulomb field - see Eq.~(\ref{29}).

Thus we obtain
\begin{equation}
\sigma_{FJ}(\omega)=\alpha\pi^2(R\delta_V(R))^2\frac{\psi_{FJ}^2(R)}{m^5}S_{FJ}(\zeta),
\label{69}
\end{equation}
with $\zeta=m/E$, and
\begin{equation}
S_{FJ}(\zeta)=S_C(\zeta),
\label{70}
\end{equation}
see Eq.~(\ref{30a}) and Fig.~\ref{fig:Fig1}.

In the ultrarelativistic limit
\begin{equation}
\sigma_{FJ}(\omega)=\alpha\pi^2(R\delta_V(R))^2\frac{\psi_{FJ}^2(R)}{m^5}\frac{m}{\omega}.
\label{71}
\end{equation}

In the same way the factors $J^2(q)$ can be obtained for the potentials which are continuous at $r=R$ while the derivative $V'(r)$ experience a jump. One should just make one more integration by parts. Two integrations by parts are needed for the potentials with continuous $V(r)$ and $V'(r)$ while $V''(r)$ experience a jump, etc.

Consider now a potential with infinite jump
\begin{equation}
V(r)= -U_0\delta(R-r);\quad U_0>0; \quad R>0,
\label{72}
\end{equation}
where $U_0$ is a dimensionless constant. It is known as the Dirac bubble potential. This potential was studied first
in~\cite{Blinder} in connection with the hyperfine splitting of the atomic levels. Its Fourier transform is
\begin{equation}
V(q)=-U_0\frac{4\pi R}{q}\sin{(qR)}.
\label{73}
\end{equation}
One can see that neither Eq.~({\ref{0b}) nor ({\ref{0c}) are true. However the integral $J(q)$ can be calculated immediately since in position space
\begin{equation}
J(q)=\int d^3re^{-i {\bf q}{\bf r}}V(r)\psi_{IJ}(r)=-4\pi U_0R\psi_{IJ}(R)\frac{\sin{(qR)}}{q},
\label{74}
\end{equation}
with $\psi_{IJ}$ the wave function of $s$ bound state in potential (\ref{72}), the lower index $IJ$ stands for "infinite jump".

To obtain the cross section we need the factor
\begin{equation}
J^2(q)=(4\pi U_0R\psi_{IJ}(R))^2\frac{\sin^2{(qR)}}{q^2}.
\label{75}
\end{equation}
Similar to the case of finite jump of the potential we present $\sin^2{(qR)}=(1-\cos{(2(qR)})/2$, with only the first term contributing to the cross section. The latter is
\begin{equation}
\sigma_{IJ}(\omega)=\alpha\pi^2(RU_0)^2\frac{\psi_{IJ}^2(R)}{m^3}S_{IJ}(\zeta),
\label{76}
\end{equation}
with
$$
S_{IJ}(\zeta)=\frac{\zeta}{8 (1-\zeta)^4}\cdot
\left(\Big(1-3\zeta+3\zeta^2-\zeta^3\Big)\ln{\frac{1+\sqrt{1-\zeta^2}}{1-\sqrt{1-\zeta^2}}}\right.
$$
\begin{equation}
\left.-2\Big(1-4\zeta+2\zeta^2\Big)\sqrt{1-\zeta^2}\right),
\label{77}
\end{equation}
with  $\zeta=m/E$, see Fig.~\ref{fig:Fig3}.

\begin{figure}
\begin{center}
\includegraphics[width=7cm]{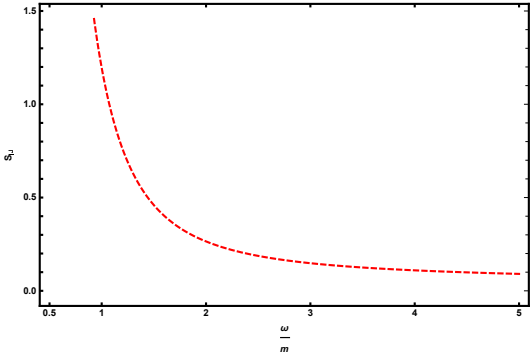}
\caption{
 Energy dependence for photoionization in the case of potential with infinite jump on the real axis defined by Eq.~(80). The horizontal axis is the same as in Fig.~\ref{fig:Fig1}. The vertical axis is for the function $S_{IJ}$ given by Eq.~(85).
}
\label{fig:Fig3}
\end{center}
\end{figure}
%Fig. 3. Energy dependence for photoionization in the case of potential with infinite jump on the real axis defined by Eq.(80). The horizontal axis is the same as in Fig.1. %The vertical axis is for the function $S_{IJ}$ given by Eq.(85).\\

In the ultrarelativistic limit $\omega \gg m$ we find
$$S_{IJ}(\zeta)=\frac{\zeta}{4}\ln{\frac{1}{\zeta}},$$
and
\begin{equation}
\sigma_{IJ}(\omega)=\alpha\pi^2(RU_0)^2\frac{\psi_{IJ}^2(R)}{4m^3}\frac{m}{\omega}\ln{\frac{\omega}{m}}.
\label{78}
\end{equation}

\section{Potentials with singularities in the complex plane}

Consider the Lorentz potential
\begin{equation}
V(r)=-\frac{U_0}{\pi}\frac{a}{r^2+a^2}; \quad a>0,
\label{42}
\end{equation}
with $U_0>0$, a dimensionless constant. One can see that $V(r)\rightarrow -U_0\delta(r)$ for $a\rightarrow 0$.
The Fourier transform is
\begin{equation}
V(q)=-\frac{2\pi U_0a}{q}\exp{(-qa)}.
\label{43}
\end{equation}
This expression can be obtained by carrying out the integration in complex plane. The potential $V(q)$ is determined by the poles of the integrand in the points $r=\pm ia$.
Since the characteristic distances for the bound state are $r \sim a$, condition expressed by Eq.~(\ref{0n1}) takes the form
\begin{equation}
ma \gg 1.
\label{43a}
\end{equation}

We need the contribution of small $f \sim 1/a \ll q$ to the integral
$$
J(q)=-2\pi U_0a\int\frac{d^3f}{(2\pi)^3}\frac{e^{-v({\bf f})a}}{v(\bf f)}\psi_B(f),
$$
\begin{equation}
v({\bf f}) =|{\bf q}-{\bf f}|= (q^2-2{\bf q}{\bf f}+f^2)^{1/2},
\label {44}
\end{equation}
which is dominated by small $f \ll q$. One can put $v=q$ in the denominator of the integrand. However this can not be done in the power of the exponential factor. To obtain the $q$ dependence of the right hand side we present Eq.~($\ref{44}$) as
$$J(q)=-2\pi\frac{U_0a}{q}X(q); \quad X(q)=\int\frac{d^3f}{(2\pi)^3}e^{-v({\bf f})a}\psi_L(f).$$
Here $\psi_L$ is the wave function of $s$ bound state in the Lorentz potential (\ref{42}).
The derivative with respect to $q$ is
$$X'(q)=-a\int\frac{d^3f}{(2\pi)^3}e^{-v({\bf f})a}v'_q({\bf f})\psi_L(f),$$
with $v_q'=\partial v/\partial q$. One can put $v'_q=1$ with the accuracy $1/q^2$. Thus we can write a simple differential equation
$X'(q)=-aX(q)$
with the solution $X(q)=\kappa e^{-qa}$ where $\kappa$ is a constant factor.
Thus
\begin{equation}
J(q)=-2\pi U_0a\kappa\frac{e^{-qa}}{q},
\label{45}
\end{equation}
and $q$ dependence of the factor $J(q)$ in the amplitude is the same as that of the potential $V(q)$.

The cross section can be obtained by employing Eqs.~(\ref{26b})-(\ref{26e})
\begin{equation}
\sigma_L(E)=4\alpha\pi^2\frac{U^2_0\kappa^2}{m^5}S_L(E),
\label{50}
\end{equation}
with
\begin{equation}
S_L(E)=e^{-2q_1a}S_L^{(1)}(\zeta)=e^{-2ma\lambda(\zeta)}S_L^{(1)}(\zeta)
\label{51}
\end{equation}
with $q_1=\sqrt{E^2-m^2}-(E-m)$ the smallest possible value of the recoil momentum $q$,
$$\lambda(\zeta)=\frac{\sqrt{1-\zeta}}{\zeta}\Big(\sqrt{1+\zeta}-\sqrt{1-\zeta)}\Big),$$
while
$$
S_L^{(1)}(\zeta)=\frac{\zeta}{8(1-\zeta)^5}\left(1+(1+\zeta-2\zeta^3)(\sqrt{1-\zeta^2}+\zeta)\right.
$$
\begin{equation}
\left. +\zeta^2(2\sqrt{1-\zeta^2}-\zeta)\right),
\label{52}
\end{equation}
see Fig.~\ref{fig:Fig4}.
Note that only the values of $q$ near its lower limit contribute since the contribution of larger values
are exponentially quenched.

\begin{figure}
\begin{center}
\includegraphics[width=7cm]{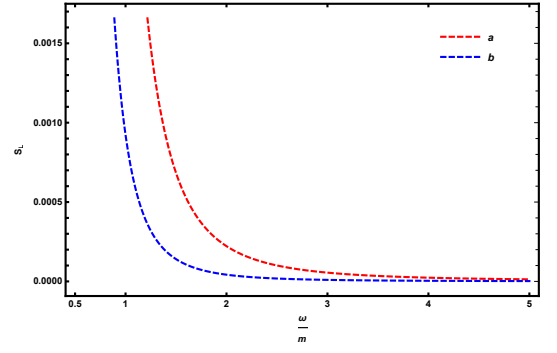}
\caption{
 Energy dependence for photoionization in the case of potential with the poles in the complex plane defined by Eq.~(87). The horizontal axis is the same as in Fig.~\ref{fig:Fig1}. The vertical axis is for the function $S_{L}$ given by Eq.~(93). Curve "a" is for $ma=5$, curve "b" is for $ma=6$.
}
\label{fig:Fig4}
\end{center}
\end{figure}
%Fig. 4. Energy dependence for photoionization in the case of potential with the poles in the complex plane defined by Eq.~(87). The horizontal axis is the same as in Fig.1. %The vertical axis is for the function $S_{L}$ given by Eq.~(93.). Curve "a" is for $ma=5$, curve "b" is for $ma=6$.\\

In the ultrarelativistic limit we find $q_1=m$, and $S_L^{(1)}(\zeta)=\zeta/8$ and
\begin{equation}
\sigma_L(\omega)=\alpha\pi^2\frac{U^2_0\kappa^2e^{-2ma}}{2m^5}\frac{m}{\omega}.
\label{53}
\end{equation}
However this behavior is achieved only at very large energies $\omega \gg m\cdot ma$ when we can replace $e^{-2q_1a}$ by $e^{-2ma}$.

It is known that the Fourier transform of the potential $V(r)$ with singularities in the complex plane experiences exponential drop $V(q) \sim e^{-qa}$ with $a$ the imaginary part of the singularity closest to the real axis~\cite{Migdal}. Thus  the cross section for relativistic photoionization of a system bound by such potential
experiences the exponential drop $\sigma \sim e^{-2q_1a}$ with $q_1=p-\omega$ the smallest possible value of the
three momentum transferred to photoelectron.

\section{Summary}

We carry out studies of relativistic photoeffect for the system bound by a central field $V(r)$. We employ
the analysis in terms of recoil momentum described in the book~\cite{Drukarev} and developed further in papers~\cite{Drukarev1,Drukarev2}.

We demonstrated that distribution in recoil momentum for relativistic photoionization in a central field $V(r)$ is expressed in terms of relativistic bound state wave function $\Psi_B(q)$  at $q \ga m$. We showed that for the potentials which satisfy the condition given by Eq.~(\ref{0b}) the wave function $\Psi_B(q)$ is connected with the  nonrelativistic wave function $\psi(f)$ at $f \sim \mu$ - Eqs.~(\ref{16b1}), (\ref{16b2}). The factor $J(q)$
given by Eq.~(\ref{16b1}) contains all dependence of the photoionization amplitude on the potential. It is expressed in terms of the  Fourier transform $V(q)$ of $V(r)$.  Since the shape of $V(q)$ depends on the structure of singularities of the potential $V(r)$ the latter determine the energy dependence of the photoionization cross section.

For the case of the potentials $V(r)$ with singularity in the origin expression for $J(q)$ takes a simple form - Eq.~(\ref{27a}). We reproduce the well known result for the Coulomb field and find the cross section for screened Coulomb potential and for the exponential potential. The recoil momentum distribution and the cross section are proportional to the nonrelativistic bound state wave function at the singular point $\psi_B(r=0)$. The results are illustrated by Figs.~\ref{fig:Fig1},~\ref{fig:Fig2}.

For the potentials with singularities on the real axis analysis of the factors $J(q)$ written in position presentation appeared to be fruitful. We demonstrate that the energy dependence for potential with a finite jump on the real axis coincides with that for the Coulomb field. The cross section depends on the squared jump of the potential explicitly, containing it as a factor. We show that the same approach can be applied for analysis of photoionization in the fields which are continuous on the real axis but have jumps of the derivatives of the first or higher order.
We consider also a potential with infinite jump on the real axis (the Dirac bubble potential) and present results in Fig.~\ref{fig:Fig3}.

We consider a potential with a pole in the complex plane and obtain the cross section presented in Fig.~\ref{fig:Fig4}. We show the cross section to be exponentially quenched. We demonstrate that this is a common feature of potentials with the poles in the complex plane.

\end{document}